\documentclass{ws-ijmpc}

\begin{document}

\markboth{Chakraborty, Jung, Khanna}
{A Multi-Domain Hybrid Method For Head-On Collision of Black-Holes in Particle Limit}

\catchline{}{}{}{}{}

\title{\textbf{A MULTI-DOMAIN HYBRID METHOD FOR HEAD-ON COLLISION OF BLACK-HOLES IN PARTICLE LIMIT}
}

\author{DEBANANDA CHAKRABORTY, JAE-HUN JUNG }

\address{Department of Mathematics, State University of New York at Buffalo,\\
Buffalo, New York, 14260-2900, United States\\
dc58@buffalo.edu, jaehun@buffalo.edu}

\author{GAURAV KHANNA}

\address{Department of Physics,  University of Massachusetts at Dartmouth,\\
Dartmouth, MA, 02747, United States\\
gkhanna@umassd.edu}

\maketitle

\begin{history}
\received{Day Month Year}
\revised{Day Month Year}
\end{history}

\begin{abstract}
A hybrid method is developed based on the spectral and finite-difference methods for solving
the inhomogeneous Zerilli equation in time-domain. The developed hybrid method decomposes
the domain into the spectral and finite-difference domains.
The singular source term is located in the spectral domain
while the solution in the region without the singular term is approximated by the higher-order finite-difference method.

The spectral domain is also split into multi-domains and the finite-difference domain is placed as the boundary domain. 
Due to the global nature of the spectral method, a multi-domain method composed of the spectral domains only  
does not yield the proper power-law decay unless the range of the computational domain is large. The finite-difference domain 
helps reduce boundary effects due to the truncation of the computational domain. The multi-domain approach with the finite-difference 
boundary domain method reduces the computational costs significantly and also yields the proper power-law decay. 

Stable and accurate interface conditions between the finite-difference and spectral domains and the spectral and spectral domains are derived. For the singular source term, we use both the Gaussian model with various values of full width at half maximum and a localized discrete $\delta$-function. The discrete $\delta$-function was generalized to adopt the Gauss-Lobatto collocation points of the spectral domain.  

The gravitational waveforms are measured.  Numerical results show that the developed hybrid method accurately yields the quasi-normal modes and the power-law decay profile. The numerical results also show that the power-law decay profile is less sensitive to the shape of the regularized $\delta$-function for the Gaussian model than expected. The Gaussian model also yields better results than the localized discrete $\delta$-function.

\keywords{Dirac $\delta$-function; finite-difference method; Spectral method; Hybrid method; Head-on collision of black holes ; Zerilli equation.}
\end{abstract}

\ccode{PACS Nos.: 02.70.Hm, 02.70.Bf, 04.30.-w}

\section{Introduction}
Since the classic works \cite{Zerilli1,Zerilli2} by Frank Zerilli in early $70$'s on the particle falling in a Schwarzschild geometry, a lot of research and study has been performed on this fundamental problem \cite{Lousto,Lousto_Price,E_Mitsou,Jung_Khanna_Nagle,Davis_Ruffini,P_Canizares,Price_Pulin,Field_Hesthaven,Nagar}. One of the earliest computational calculations was made by Press and his co-workers, which is now known as DRPP calculation \cite{Davis_Ruffini} on the radiation emitted when a particle starting from rest at infinity falls into a non-spinning black hole. The collision of two black holes is, in principle, one of the most efficient mechanisms for the generation of the gravitational waves. In recent years the extreme mass ratio limit of the binary system has been a special focus of research in gravitational physics. Extreme-Mass-Ratio Inspirals (EMRIs) are one of the main sources of the gravitational waves for the gravitational wave detectors, such as Laser Interferometer Space Antenna (LISA) \cite{Lisa}. EMRIs are binary systems composed of a stellar compact object (SCO) with a mass, $m$ in the range of $m = 1  \sim 10^2M_{\bigodot}$ inspiralling into a massive black hole (MBH) with a mass, $M$ in the range of $M = 10^4  \sim  10^7M_{\bigodot}$ located at the galactic center. Thus, the mass ratios involved are $\mu := m/M \sim 10^{-7}-10^{-2}$. During the slow inspiral phase the system is driven by the emission of gravitational radiation, the general features of which are now well understood. Press showed that there is always an intermediate stage where the ringdown is dominated by a set of oscillating and exponentially decaying solutions, quasinormal modes (QNMs) whose spectrum depends only on the mass of the black hole and the multipole-moment index $l$ of the initial perturbation \cite{Pazos_Brizuela}. This regime is followed by a power-law {\it tail} decay due to backscattering.

For the EMRI, the small companion black hole is modeled as a point particle, and the problem can be framed by using the black hole perturbation theory. Moreover, as the first approximation, the point particle follows the geodesics in the space-time of the central black hole. The frequency-domain approach to this problem has achieved many remarkable results. Specifically the accurate $(\lesssim 10^{-4})$ determination of the energy flux of gravitational waves was obtained in the frequency-domain \cite{Glampedakis}. However, the frequency-domain approach can take long computational time and lose accuracy for non-periodic orbits (for example, parabolic orbits,  orbits with high eccentricity or decaying orbits). The time-domain approach seems better suited for such orbits \cite{Burko_Khanna}. For the time-domain approach, the finite-difference (FD) method is  one of the most popular numerical methods. The FD time-domain methods, however, suffer from the relatively poor accuracy at the moment \cite{Sundararajan_Khanna_Hughes} unless a very high computational resolution is used. The main reason is the point particle approximation, i.e. the approximation of the singular source terms. Various approaches to this issue have been attempted, including the regularizing the Dirac $\delta$-function using a narrow Gaussian distribution and also using more advanced discrete $\delta$-model \cite{Jung_Khanna_Nagle,Gottlieb_Orszag,Sundararajan_Khanna_Hughes,Sundararajan_Khanna_Hughes_Drasco}.

Another approach of the EMRI problem is to use the spectral (SP) method \cite{Jung_Khanna_Nagle,Jung,Hesthaven_Gottlieb_Gottlieb,P_Canizares}. In our previous work, we used the spectral method to solve the inhomogeneous Zerilli equation in time-domain and obtained good results. But the proper power-law decay was not observed \cite{Jung_Khanna_Nagle}. In early time the solution agrees with the established solution but in very late time the solution is contaminated by the  small-scale oscillations. These oscillations are likely due to the artificial truncation of the computational domain. 

In this work, we continue our previous research with the spectral method in order to obtain the proper power-law decay. For this, we developed the multi-domain hybrid method. The multi-domain method hybridizes the spectral method and the high-order finite-difference method. The spectral domain is also split into many sub-domains, each of which is also a spectral domain. The main advantage of the multi-domain method is that the computational costs can be significantly reduced by reducing the order of the interpolating polynomial in each sub-domain and the parallelization becomes robust. A fundamental reason for considering the multi-domain method is also to reduce the boundary effects due to the artificial truncation of the computational domain for obtaining the proper late time decay profile of the gravitational waveforms. In order to obtain the proper power-law decay, the outer boundary needs to be placed afar, in general. However, having a large size of the computational domain increases the computational costs significantly. In this work, we add the finite-difference domain as the boundary domain. The spectral method is a global method and it is highly sensitive to the boundary effects. To prevent the ``fast'' propagation of these boundary effects, we use a local method instead as the boundary domain, such as the finite-difference domain. By doing this, we obtain the proper power-law decay while having the computational costs reduced and also exploiting the accuracy of the spectral method. To patch each sub-domain with others, we derive the accurate and stable patching conditions. For the spectral and finite-difference sub-domains, we show that the resolution across the interface needs to be closely uniform. Otherwise, the CFL condition becomes strict. 

For the singular source term, we use both the Gaussian $\delta$-function method and the discrete $\delta$-function method. For the Gaussian method, we change the shape of the Gaussian profile to mimic the $\delta$-function.  For the discrete $\delta$-function, we generalize the discrete $\delta$-function developed by Sundararajan et al. \cite{Sundararajan_Khanna_Hughes,Sundararajan_Khanna_Hughes_Drasco} into the one on the non-uniform grid. 

We provide numerical results that show the efficiency and robustness of the proposed hybrid method. Using the hybrid method we could obtain the proper power-law decay with the Gaussian approximation model. We use various shapes of the Gaussian profile and found that the result is insensitive to the shape. That is, even a broad profile, which results in a smooth solution, yields the power-law decay successfully. With the smooth solution, the spectral method does not need to use the filter operation, which increases the computational efficiency further. We also obtain the power-law decay with the discrete $\delta$-function model, but the computed slope was not accurate, which may imply that the discrete $\delta$-function model yields correct results only on uniform grids.

This paper is organized as follows. In Section 2,  we briefly describe the finite-difference and spectral methods. For the finite-difference method, we used the $4$th-order  method.  For the spectral method we use the Chebyshev spectral collocation method based on the Gauss-Lobatto collocation points. In Section 3  we describe the discrete $\delta$-function on non-uniform grids. Section 4 explains the Zerilli equation briefly. In Section 5, we describe the proposed hybrid method in details. We derive the stable and accurate interface patching conditions. Boundary conditions are described in Section 6. In Section 7, we discuss the stability of the hybrid method. In Section 8, numerical results are provided. In Section 9, a brief summary and future work are explained.

 
\section{finite-difference and spectral methods}

\subsection{finite-difference method}
In this work we consider both the 2nd and 4th-order finite-difference method. The 2nd-order finite-difference method for the spatial 
derivatives are well known and we omit those formulae. Instead we briefly explain the 4th-order finite-difference method. For the 4th-order method we will derive the formula when the grid is non-uniform. This is because, in the spectral domain, we use the Gauss-Lobato collocation points, which are not evenly distributed and we need the finite-difference formula for the boundary conditions in the spectral domain. Also we define the modified flux at the SP-SP interface, which also requires the finite-difference formulae. Details of these are described in section $5$.

{\em 4th-Order finite-difference method: Uniform grids.} 
Let $h$ be the grid spacing in the finite-difference domain and let $\xi$ be $x \le \xi \le x + h$. 
The standard $4th$-order derivatives are given by 

Centered difference:
\begin{eqnarray}
\quad  u^{\prime}(x)|_{x = x_{j}} = \frac{u_{j-2}-8u_{j-1}+8u_{j+1}-u_{j+2}}{12h}-\frac{1}{30}h^{4}u^{(5)}(\xi).
\end{eqnarray}

Off-centered (1 - point) difference: 
\begin{eqnarray}
u^{\prime}(x)|_{x = x_{j}} = \frac{-3u_{j-1}-10u_{j}+18u_{j+1}-6u_{j+2}+u_{j+3}}{12h}+\frac{1}{20}h^{4}u^{(5)}(\xi).
\end{eqnarray}

Off-centered (2 - points) difference:
\begin{eqnarray}
u^{\prime}(x)|_{x = x_{j}} = \frac{-25u_{j}+48u_{j+1}-36u_{j+2}+16u_{j+3}-3u_{j+4}}{12h}-\frac{1}{5}h^{4}u^{(5)}(\xi).
\end{eqnarray}

At the left $x = x^-$ and right $x = x^+$ boundaries, we used the 2nd-order difference method.

$x = x^-$: $x^- < \xi < x^- + h$
\begin{eqnarray}
\label{left2ndorder}
u^{\prime}(x^-) = \frac{-3u(x^-)+4u(x^-+h)-u(x^-+2h)}{2h}+\frac{2}{3}h^{2}u^{(3)}(\xi),\\
\; where \; x^{-} < \xi < x^{-}+h. \nonumber
\end{eqnarray}
Similarly 

$x = x^+$: $x^+ - h < \xi < x^+$ 
\begin{eqnarray}
\label{right2ndorder}
u^{\prime}(x^+) =  \frac{3u(x^+)-4u(x^+-h)+u(x^+-2h)}{2h}+\frac{2}{3}h^{2}u^{(3)}(\xi),\\
\; where \; x^{+}-h < \xi < x^{+}. \nonumber
\end{eqnarray}
The centered 2nd-order derivative is given by 
  \begin{equation} 
  \label{equation19}
  u^{\prime \prime}(x) = \frac{-u_{j-2}+16u_{j-1}-30u_{j}+16u_{j+1}-u_{j+2}}{12h^2}-\frac{1}{90}h^4u^{(6)}(\xi).
  \end{equation}

\subsection{Spectral method}
For the spectral method, we use the Chebyshev spectral collocation method based on the Gauss-Lobatto collocation points.  
The Chebyshev spectral collocation method seeks a solution in the Chebyshev polynomial space by the Chebyshev polynomials $T_{l}(x)$ as 
$$ u_{N}(x,t)= P_{N}u(x,t) = \sum_{l=0}^{N}\widehat{u}_{l}(t)T_{l}(x),$$ 
where $ T_{l}(x)$ is the Chebyshev polynomial of degree $ l $ and $\widehat{u}_{l}$ the corresponding expansion coefficient. The commonly used collocation points are the Gauss-Lobatto quadrature points $x_{j}$  given by $$ x_{j} = \cos(\frac{\pi}{N}j),\quad \forall j = 0,\cdots, N.$$ These collocation points belong to $[-1\:1]$ and so we need the proper mapping to fit in our computational domain. The required mapping is $$X_{j} = a + \frac{b-a}{2}(x_{j}+1),$$ where $[a ,\:b]$ is the original computational domain.
The expansion coefficients are given by  $$\widehat{u}_{l}(t) = \frac{2}{c_{l}N}\sum_{j=0}^{N}\frac{1}{c_{j}}u_{j}(t)T_{l}(x_{j}),$$  where $c_{n} = 2$ if $n = 0$ , and $c_{n}=1$ otherwise.

\subsection{Spectral filtering method}

We also use the spectral filtering method to minimize the possible non-physical high frequency modes. The oscillations with the spectral method possibly found near the local jump discontinuity and also generated due to inconsistent initial conditions propagate through the whole domain. Our filtered approximation is given by $$u^{\sigma}_{N}(x,t)=\sum_{j=0}^{N}\sigma(j/N)\widehat{u}_{j}T_{j}(x),$$ where $\sigma(l/N)$ based on the exponential filter is the filter function according to \cite{Gottlieb_Orszag,Hesthaven_Gottlieb_Gottlieb}. Our filter matrix $S$ is given by $$S_{ij}=\frac{1}{c_{j}}\sum_{n=0}^{N}\frac{2}{c_{n}N}T_{n}(x_{i})T_{n}(x_{j})\exp(ln\epsilon(n/N)^{p}),$$ where $p$ is the order of filteration and $\epsilon$ is a constant. The filtered solution at $x = x_{i}$ is given by  $$u^{\sigma}_{N}(x_{i})=\sum_{j=0}^{N}S_{ij}u_{j}, \; \forall i = 0,\cdots ,N.$$

\section{Approximation of singular source}

\subsection{Discrete $\delta$-function with non-uniform grid}
Discrete $\delta$-function on a uniform grid has been derived in \cite{Sundararajan_Khanna_Hughes_Drasco}. In SP-FD approach, the singular source term is always located inside the spectral domain. Since in the spectral domain the grid is non-uniform \cite{Hesthaven_Gottlieb_Gottlieb}, we need to redefine the discrete $\delta$-function on the non-uniform grid.
$\delta$-function which exists at $x = \alpha$,  $x_{k+1}\le \alpha \le x_{k+2}$, by using the following relation
 \begin{eqnarray}
 \sum_{i} h_{i}f(x_{i})\delta_{i} &=&  f(\alpha) \nonumber \\ 
                                          &= & \frac{(\alpha - x_{k+1})(\alpha - x_{k+2})(\alpha - x_{k+3})}{(x_{k} - x_{k+1})(x_{k} - x_{k+2})(x_{k} - x_{k+3})}f(x_{k})\nonumber \\
                                          && + \frac{(\alpha - x_{k})(\alpha - x_{k+2})(\alpha - x_{k+3})}{(x_{k+1} - x_{k})(x_{k+1} - x_{k+2})(x_{k+1} - x_{k+3})}f(x_{k+1}) \nonumber \\ 
                                          && + \frac{(\alpha - x_{k})(\alpha - x_{k+1})(\alpha - x_{k+3})}{(x_{k+2} - x_{k})(x_{k+2} - x_{k+1})(x_{k+2} - x_{k+3})}f(x_{k+2}) \nonumber \\
                                           &&+ \frac{(\alpha - x_{k})(\alpha - x_{k+1})(\alpha - x_{k+2})}{(x_{k+3} - x_{k})(x_{k+3} - x_{k+1})(x_{k+3} - x_{k+2})}f(x_{k+3}). \nonumber 
\end{eqnarray}
 Equating the coefficients of $f(x_{i})$ from both sides yields 
 \begin{equation} 
\delta_{i} = \left\{ 
 \begin{array}{lr}
   -\frac{(\alpha-x_{k+1})(\alpha - x_{k+2})(\alpha - x_{k+3})}{(x_{k+1} - x_{k})^{2}(x_{k+2} - x_{k})(x_{k+3} - x_{k})}     & \mbox{ at }  x_{k}\\
    \frac{(\alpha-x_{k})(\alpha - x_{k+2})(\alpha - x_{k+3})}{(x_{k+1} - x_{k})(x_{k+2} - x_{k+1})^{2}(x_{k+3} - x_{k+1})}  & \mbox{ at } x_{k+1} \\
    -\frac{(\alpha-x_{k})(\alpha - x_{k+1})(\alpha - x_{k+3})}{(x_{k+2} - x_{k})(x_{k+2} - x_{k+1})(x_{k+3} - x_{k+2})^{2}}  &
   \mbox{ at } x_{k+2} \\
    \frac{(\alpha-x_{k})(\alpha - x_{k+1})(\alpha - x_{k+2})}{(x_{k+3} - x_{k})(x_{k+3} - x_{k+1})(x_{k+3} - x_{k+2})(x_{k+4}-x_{k+3})} &\mbox{ at } x_{k+3}.
\end{array}\right.
\end{equation}

For getting the first derivative of the $\delta$ function we have 
\begin{eqnarray}
 \sum h_{i} f(x_{i}) \delta_{i}^{'} &=& -f^{'}(\alpha) \nonumber \\
                                                 & = &  -\sum h_{i}f^{'}(x_{i})\delta_{i} \nonumber  \\
                                                 & = &  -\sum h_{i}\delta_{i}\frac{f(x_{i+1})-f(x_{i-1})}{(x_{i+1}-x_{i-1})}.\nonumber  
\end{eqnarray}
 Equating the coefficients of $f(x_{i})$ from both sides yields 
\begin{equation} 
 {\delta^{'}_i} = \left\{ 
\begin{array}{ll}
 \frac{(x_{k+1}-\alpha)(\alpha - x_{k+2})(\alpha - x_{k+3})}{(x_{k}-x_{k-1})(x_{k+1}-x_{k-1})(x_{k+1}-x_{k})(x_{k+2}-x_{k})(x_{k+3}-x_{k})} &
    \mbox{ at }  x_{k-1}\\
 \\
 \frac{(\alpha-x_{k})(\alpha - x_{k+2})(\alpha - x_{k+3})}{(x_{k+1}-x_{k})^{2}(x_{k+2}-x_{k})(x_{k+2}-x_{k+1})(x_{k+3}-x_{k+1})} &
    \mbox{ at } x_{k} \\
\\
%
  \frac{(\alpha-x_{k+1})(\alpha - x_{k+3})}{(x_{k+2}-x_{k+1})(x_{k+2}-x_{k})} 
  [\frac{\alpha-x_{k+2}}{(x_{k+1}-x_{k-1})(x_{k+1}-x_{k})(x_{k+3}-x_{k})}  & \mbox{ at } x_{k+1} \\
 \quad\quad -\frac{\alpha-x_{k}}{(x_{k+3}-x_{k+1})(x_{k+2}-x_{k+1})(x_{k+3}-x_{k+2})}]  
  & 
\\
\\
  \frac{(\alpha-x_{k})(\alpha - x_{k+2})}{(x_{k+3}-x_{k+2})(x_{k+3}-x_{k+1})}[\frac{\alpha-x_{k+1}}{(x_{k+4}-x_{k+2})(x_{k+3}-x_{k})(x_{k+3}-x_{k+2})}
  & \mbox{ at }  x_{k+2}\\
  \quad\quad -\frac{\alpha-x_{k+3}}{(x_{k+2}-x_{k})(x_{k+1}-x_{k})(x_{k+2}-x_{k+1})}] & 
\\
\\
  \frac{(\alpha - x_{k})(\alpha - x_{k+1})(\alpha - x_{k+3})}{(x_{k+3}-x_{k+1})(x_{k+2}-x_{k})(x_{k+2}-x_{k+1})(x_{k+3}-x_{k+2})(x_{k+4}-x_{k+3})}
 & \mbox{ at } x_{k+3} \\
 \\
   -\frac{(\alpha - x_{k})(\alpha - x_{k+1})(\alpha - x_{k+2})}{(x_{k+5}-x_{k+4})(x_{k+4}-x_{k+2})(x_{k+3}-x_{k})(x_{k+3}-x_{k+1})(x_{k+3}-x_{k+2})}
   &\mbox{ at } x_{k+4}. 
\end{array}
\right. 
\end{equation}

\subsection{Gaussian $\delta$-function}
The Gaussian $\delta$-function is defined by 
$$\delta(x) = \frac{1}{\sqrt{\pi\sigma}}e^{-\frac{(x-\mu)^2}{\sigma}}.$$ The shape of the $\delta$-function  depends on the full width at half maximum $\sigma$. If $\sigma$ is small, $\delta(x)$ is narrow and if $\sigma$ has higher value then $\delta(x)$ is broader. The value of $\mu$ depends on time and determines the position of the $\delta$-function.

\section{Zerilli equation}
The lowest order perturbation theory of the initial Schwarzschild black hole spacetime leads to the inhomogeneous Zerilli equation with even-parity \cite{Zerilli1}. Such an equation describes the gravitational wave $\psi$ in $1 \, + \, 1$ dimension given by the following second-order wave equation 
\begin{eqnarray}
\psi_{tt} = \psi_{r^{*}r^{*}} - V_{l}(r)\psi - S_{l}(r,t),
\end{eqnarray}
where $r^{*}$ is the tortoise coordinate, $V_{l}(r)$ the potential term and $S_{l}(r,t)$ the source term. For details see [3,10]. The tortoise coordinate, is given by 
\begin{eqnarray}
r^{*} = r + 2Mln(r/2M-1),
\end{eqnarray} where $r$ is the physical coordinate and $r > 2M$.

One can convert Eq (9) into a system of equations. In some previous works, Zerilli equation was solved in time-domain \cite{Pazos_Brizuela,P_Canizares} and in all cases the authors converted the $2$nd-order PDE to the system of $1$st-order equations in space and time. We understand that some instabilities arose and to suppress those instabilities, they introduced an auxiliary field which converted the equation to a coupled set of $1$st-order equations in space and time.

In this work we do not convert the $2$nd-order PDE to the system of equations. Instead we control the instability efficiently by introducing new interface conditions between SP-SP interfaces and SP-FD interfaces. We also correct the numerical flux accordingly.

\section{Hybrid method} 
The hybrid method is basically the domain decomposition method. We use the finite-difference and the spectral methods for the hybrid method. The one-dimensional single domain is decomposed into multi-domains. Each domain is carried by either the finite-difference method or the spectral method. 

Let $\Omega = [R_e, R_\infty]$ be the original domain. Suppose that $\Omega$ is decomposed into $M$ sub-domains 
and each sub-domain meets its adjacent sub-domain at its boundaries only 
such that 
$$
     \Omega = \bigcup_{i=1}^M \Omega^i, \quad      \Omega^i \bigcap \Omega^j = \partial^+ \Omega^i, \mbox{ or } \partial^- \Omega^j, 
     \quad j = i + 1,
$$
where $\Omega^i$ is the $i$th sub-domain and $\partial^{+}\Omega$ and $\partial^{-}\Omega$ denote the right and left boundaries of the sub-domain.  
Let $\Omega^i_s$ and $\Omega^i_f$ denote 
the $i$th sub-domain of the SP domain and the FD domain respectively. 
The $i$th sub-domain $\Omega^i$ should be either $\Omega^i_s$ or $\Omega^i_f$. 
Let $N^i + 1$ be the total number of grid points of $\Omega^i$
and let $\{x^i_0, \cdots, x^i_{N^i}\}$ be the set of grid points in $\Omega^i$. 
The sub-domains are in order, indexed by the index $i$ from $1$ to $M$. That is, $i = 1$ is corresponding  to the leftmost sub-domain and 
$i=M$ the rightmost sub-domain. In this work, by default, we let the last domain be the FD domain, i.e. $\Omega^M = \Omega^M_f$. The reason is to control the boundary effects arising from the outer boundary. We found some small oscillations arise from the outer boundary that contaminate the power-law decay when the multi-domain spectral method was used in late times. Since the FD method is the local method, we put a FD domain as the boundary domain to remove the boundary effect.
   
The main development of the hybrid method is the stable patching algorithm at the domain interfaces between the SP and FD domains and the SP-SP domains. This involves the approximation of the derivatives across the non-uniform grid points, for which we consider the 4th-order FD method with the non-uniform  grid and the polynomial interpolation method. 

\subsection{4th-order finite difference method: non-uniform grids}
The 4th-order centered FD method uses 
$5$ points and the non-uniformity needs to be addressed in the FD approximation of the derivatives
while the 2nd-order centered FD method does not need such non-uniform adaptation. 
The schematic diagram is given in Diagram 2. 

{\em 2nd-order finite-difference derivative at the spectral domain boundary:}
FD domain imposes the outflow boundary condition $(\partial_t + \partial_x)u = 0$ at the right domain boundary. 
In the case that the last domain is a SP domain, we adopt the non-uniform FD method 
for the term of $\partial_x$ in the outflow boundary condition.

Let $h_i = x_i - x_{i-1}$ and $h_{i-1} = x_{i-1} - x_{i-2}$ and assume that $h_i$ and $h_{i-1}$ are all non-zero constants and are not necessarily the same. 
With these definitions, let 
$$
    \lambda = \frac{h_{N_{M}-1}}{h_{N_{M}}}.                       
$$ 
Hereafter let {\it end} and {\it ghost} denote the indices of the last grid points and the ghost grid points respectively. 
Then the derivative $U'(x)$ at $x = x^M_{N_M}$ for $\Omega^M_s$ is given by
%
\begin{equation}
\label{right2ndorder2}
U^{\prime}(x) \approx \frac{(\lambda^2-1)U_{N_M}-\lambda^2 U_{N_M-1}+U_{N_M-2}}{\lambda(\lambda-1)h_{end}}.
\end{equation}
 
 {\em 4th-order finite-difference derivative at the interface across the adjacent spectral domains:}
At the domain interfaces across the SP domains, we use the 4th-order
FD method for the 2nd-order derivative (see Diagram 1). 
Let 
$x_{j}-x_{j-1} = x_{j+1}-x_{j} = h_{j}$ , $x_{j}-x_{j-2} = x_{j+2}-x_{j} = h_{j+1}$ and 
$$
\lambda = \frac{h_{j+1}}{h_{j}}.
$$
Then we have the 2nd-order derivative at $x = x_j$
\begin{equation} 
U^{\prime\prime}(x_{j})\approx \frac{-U_{j-2}+\lambda^4 U_{j-1}-2(\lambda^4-1)U_{j}+\lambda^4 U_{j+1}-U_{j+2}}{h_{j}^{2}\lambda^2(\lambda^2-1)}. 
\end{equation}

Using the above relation, the derivative at $x = x^i_0$ for the SP domain $\Omega^i_s$ is given by 
\begin{equation} 
\label{equation20}
U^{\prime\prime}(x_{0}^{i})\approx \frac{-U^{i-1}_{N_{i-1}-2}+\lambda^4 U^{i-1}_{N_{i-1}-1}-2(\lambda^4-1)U^i_{0}+\lambda^4 U^i_{1}-U^i_{2}}{h_{j}^{2}\lambda^2(\lambda^2-1)}. 
\end{equation}

\subsection{High-order interpolation}
For the interface patching between the FD domain and the SP domain, we use the high-order interpolation based on the Lagrange 
interpolation. For the interpolation at the ghost cells in the FD domain for the SP approximation in $\Omega^i_s$, 
we use the $4$th-order polynomial. For example, for FD-SP patching we use the $5$ points in the FD domain, 
$\Omega^{i-1}_f$.  These points are $z_k = x^{i-1}_{N^{i-1} - k}$, $k = 0, 1, 2, 3, 4$. The points for the ghost cells to seek are $y_1$ and $y_2$ (two grid points marked with the symbol $\times$ in Diagram 2) and  
these are satisfying the following relations
\begin{eqnarray}
                           && y_1 < y_2 < x^{i-1}_{N^{i-1}}, \nonumber \\
                           && x^{i-1}_{N^{i-1}} - y_2 = x^i_1 - x^i_0, \nonumber \\
                           && y_2 - y_1 = x^i_2 - x^i_1. \nonumber
\end{eqnarray}
We use the Lagrange interpolation to find the interpolation $U(y)$ at $y = y_1$, and $y = y_2$ such that
\begin{equation}
\label{FDinterpolation}
                 U(y_j) = \sum_{k=0}^4 U^{i-1}(z_k)\prod_{l = 0, l \ne k}^4 {{y_j - z_l}\over{z_k - z_l}},\quad j = 1, 2. 
\end{equation}

For the interpolation at the ghost cells in the SP domain for the FD domain  ($\Omega^i_f$, e.g.) we use the 
Chebyshev interpolation in the spectral domain ($\Omega^{i+1}_s$). The grid points for the ghost cells in the SP domain 
are $w_1$ and $w_2$ and these are satisfying the following relations (see Diagram 2)
\begin{eqnarray}
                      && x^{i+1}_0 < w_1 < w_2, \nonumber \\
                      && w_1 - x^{i+1}_0 = x^i_{N^i} - x^i_{N^i-1}, \nonumber \\
                      && w_2 - w_1 = x^i_{N^i-1} - x^i_{N^i-2}. \nonumber
\end{eqnarray}
Since the FD domain has the uniform grids, we have $ w_1 - x^{i+1}_0 =  w_2 - w_1 = x^i_{N^i} - x^i_{N^i-1}$. Similarly to the 
FD interpolation, the interpolation $U(w)$ at $w = w_1$, and $w = w_2$ is given by for $j = 1,2$
\begin{equation}
\label{SPinterpolation}
            U(w_j) = \sum_{k=0}^{N^{i+1}} U(x^{i+1}_k) L_k(\xi(w_j)), \quad j = 1, 2. 
\end{equation}
Note that $\xi$ is the linear mapping from $[x^{i+1}_{0},x^{i+1}_{N^{i+1}}]$ to $[-1 \quad 1]$. 
Here $L_k(w)$ is the Lagrange interpolation polynomial based on the Chebyshev polynomials given by 
$$
                L_k(w) = {{(-1)^{N^i+1}(1-w^2)T'_{N^{i+1}}(w)}\over{{\bar c}_k{N^{i+1}}^2(w - w_k)}}, 
$$
where ${\bar c}_j = 2$ if $j = 0, N^{i+1}$ and ${\bar c}_j = 1$ otherwise \cite{Hesthaven_Gottlieb_Gottlieb} and $T'_{N^{i+1}}(w)$ is the
derivative of the Chebyshev polynomial of degree $N^{i+1}$ with respect to $w$. Since the polynomial order in each SP domain is low, the interpolation can be done quickly without using the fast transformation. 

\subsection{Patching procedure} 
To explain the patching procedure, we consider the following 2nd-order wave equation, $u_{tt} - u_{xx} = 0$. 
 Including the potential term is straightforward.


{\underline {\bf SP-SP patching}:}
Consider the two adjacent SP domains $\Omega^{i+1}_s$ and $\Omega^i_s$ whose solutions are denoted by 
$\phi$ and $\psi$, respectively. (Hereafter, let $\phi$ and $\psi$ denote the solution in the left and the right sub-domains, respectively.)
The patching is based on Eq. (\ref{equation20}) at the domain boundaries only. For $\Omega^{i+1}_s$
the left boundary value is updated based on the following $4$th-order method: 

\begin{equation}
\label{equation25}
\phi_0^{j+1} = 2 \phi^j_0 - \phi^{j-1}_0 + \Delta t^2
\frac{-\psi^j_{N^{i}-2} + \lambda^4 \psi^j_{N^{i}-1}
-2(\lambda^4-1)\phi^j_{0} + \lambda^4 \phi^j_{1}-\phi^j_{2}}{h_{o}^{2}\lambda^2(\lambda^2-1)},
\end{equation}


where the superscripts $j+1$, $j$ and $j-1$ indicate the time steps $(j+1)\Delta t$, $j\Delta t$, and $(j-1)\Delta t$ and 
$h_o = x^{i+1}_1 - x^{i+1}_0$ and $\lambda = (x^{i+1}_2 - x^{i+1}_0)/h_o$. Note that we let every SP domain has the 
same domain interval. Thus $h_o = x^i_{N^i} - x^i_{N^i-1}$ and $x^{i+1}_2 - x^{i+1}_0 = x^{i}_{N^i} - x^i_{N^i-2}$. 

\vskip 0.5in
\setlength{\unitlength}{4cm}
\begin{picture}(1,1)
\put(1.25,1.25){\vector(-1,-1){0.2}}
\put(0.75,1){\circle{0.09}}
\put(0.9,1){\circle{0.09}}
\put(1.1,1){\circle{0.09}}
\put(1.25,1){\circle{0.09}}

\put(0.1,1){\circle*{0.05}}
\put(0.5,1){\circle*{0.05}}
\put(0.75,1){\circle*{0.05}}
\put(0.9,1){\circle*{0.05}}
\put(1.0,1){\circle*{0.05}}
\put(1.0,1){\circle*{0.05}}
\put(1.1,1){\circle*{0.05}}
\put(1.25,1){\circle*{0.05}}
\put(1.5,1){\circle*{0.05}}
\put(1.9,1){\circle*{0.05}}

\put(1.27,1.25){$\partial\Omega^i_s, \partial\Omega^{i+1}_s$}

\put(-0.1,1){\line(5,0){2.3}}
\put(0,0.5){$\begin{array}{clrl}
{\bullet}  &\mbox{Spectral grids,}    & \bigcirc & \mbox{Spectral ghost cells}\end{array}$ }
\put(-0.1,0.3){Diagram 1: SP - SP patching. Grids at the SP and SP interface
}
\label{Diagram1}
\end{picture}

For $\Omega^i_s$ the right boundary value is updated in the same way:
\begin{equation}
\label{equation26}
\psi^{j+1}_{N^i} =2 \psi^{j}_{N^i} - \psi^{j-1}_{N^i} + 
\Delta t^2 \frac{-\psi^j_{N^{i}-2} + \lambda^4 \psi^j_{N^{i}-1}
-2(\lambda^4-1)\psi^j_{N^i} + \lambda^4 \phi^j_{1}-\phi^j_{2}}{h_{o}^{2}\lambda^2(\lambda^2-1)}. 
\end{equation}

\vskip 1in
\setlength{\unitlength}{4cm}
\begin{picture}(1,1)
\put(-0.2,1){\circle{0.09}}
\put(0.2,1){\circle{0.09}}
\put(0.6,1){\circle{0.09}}
\put(1.0,1){\circle{0.09}}
\put(1.0,1){\circle*{0.05}}
\put(1.1,1){\circle*{0.05}}
\put(1.25,1){\circle*{0.05}}
\put(1.5,1){\circle*{0.05}}
\put(1.9,1){\circle*{0.05}}
\put(2.5,1){\circle*{0.05}}

\put(1.25,1.25){\vector(-1,-1){0.2}}
\put(1.27,1.25){$\partial\Omega^i_f, \partial \Omega^{i+1}_s$}

\put(0.9,0.98){$\times$}
\put(0.75,0.98){$\times$}

\put(1.4,0.98){$\bigtriangleup$}
\put(1.8,0.98){$\bigtriangleup$}

\put(-0.3,1){\line(7,0){3}}
\put(0,0.5){$\begin{array}{clrl}
\bigcirc &\mbox{finite-difference grids,}& \bigtriangleup &\mbox{finite-difference ghost cells}\\
{\bullet}  &\mbox{Spectral grids,}    & \times & \mbox{Spectral ghost cells} \end{array}$ }
\put(-0.1,0.3){Diagram 2: FD - SP patching. Grids at the FD and SP interface.
}
\label{diagram2}
\end{picture}

{\underline{\bf FD-SP patching}:}
For the FD and SP domain patching, we use the polynomial interpolation for the ghost cells. For this case, the $4$th-order
method is used based on  Eq. 6 for the uniform grid. Let $\psi$ and $\phi$ be the solutions in the
FD domain $\Omega^i_f$ and SP domain $\Omega^{i+1}_s$, respectively. The value at the right 
boundary of $\Omega^i_f$ is updated using the following:
  \begin{equation}
  \label{equation27} 
  \psi^{j+1}_{N^i} = 2 \psi^{j}_{N^i} - \psi^{j-1}_{N^i} + 
\Delta t^2 \frac{-\psi^j_{N^{i}-2} + 16 \psi^j_{N^{i}-1}
-30\psi^j_{N^i} + 16 \Psi^j_{1}-\Psi^j_{2}} {12 h_f^2},
  \end{equation}
  where $h_f = x^i_{N^i} - x^i_{N^i-1}$ and $\Psi^j_l$ are from Eq. (\ref{SPinterpolation})
$$
            \Psi^j_l = \sum_{k=0}^{N^{i+1}} U(x^{i+1}_k) L_k(\xi(x^{i+1}_{N^{i+1}} + l h_f)), \quad l = 1, 2. 
$$

The value at the left boundary of $\Omega^{i+1}_s$ is updated in the similar way, but using the non-uniform formula Eq. (\ref{equation25}), we have
\begin{equation}
\label{equation28}
\phi_0^{j+1} = 2 \phi^j_0 - \phi^{j-1}_0 + \Delta t^2
\frac{-\Phi^j_{1} + \lambda^4 \Phi^j_{0}
-2(\lambda^4-1)\phi^j_{0} + \lambda^4 \phi^j_{1}-\phi^j_{2}}{h_{o}^{2}\lambda^2(\lambda^2-1)},
\end{equation}
where $h_o$ and $\lambda$ are defined in the same way as in Eq. (\ref{equation25}). And the interpolation values $\Phi^j_l$ are from Eq. (\ref{FDinterpolation})
given by 
$$
    \Phi^j_m = \sum_{k=0}^4 U^{i}(z_k)\prod_{l = 0, l \ne k}^4 {{y_m - z_l}\over{z_k - z_l}},\quad m = 0, 1, 
$$
where $z_k = x^i_{N^i - k}$ and 
$$
y_m = x^i_{N^i} - h_o \lambda^m, \quad m = 0,1.
$$

{\underline{\bf Flux correction}}: 
After the patching procedure for the FD-SP and SP-SP patching, we consider the flux correction. Since we set each SP domain to have the same resolution, it is obvious that
$$
\phi^{j}(ghost) = \phi^{j}(2),
$$
$$
\psi^{j}(ghost) = \psi^{j}(end - 1).
$$
We use the central (average) flux to get 
\begin{eqnarray}
Flux &=& \frac{1}{2} [\psi^{j+1}(end)+\phi^{j+1}(1)]\nonumber \\
&=&\frac{1}{2}[2\psi^{j}(end)-\psi^{j-1}(end)+2\phi^{j}(1)-\phi^{j-1}(1)] \nonumber \\
&+&\frac{1}{2}(\frac{\Delta t}{h})^{2}[2\phi^{j}(2)+2\psi^{j}(end-1)-2\psi^{j}(end)-2\phi{j}(1)] \nonumber \\
&=&\frac{1}{2}[2\psi^{j}(end)-\psi^{j-1}(end)+2\phi^{j}(1)-\phi^{j-1}(1)]+\nonumber \\
&&\frac{(\Delta t)^2}{h}[\frac{\phi^{j}(2)-\phi{j}(1)}{h}-\frac{\psi^{j}(end)-\psi^{j}(end-1)}{h}] \nonumber \\
&=&\frac{1}{2}[2\psi^{j}(end)-\psi^{j-1}(end)+2\phi^{j}(1)-\phi^{j-1}(1)]+\frac{(\Delta t)^2}{h}[\phi_{x}(1)-\psi_{x}(end)].\nonumber
\end{eqnarray}

If the solution is smooth, e.g. $\psi, \phi \in C^{1}$ then $\phi_{x}(1)=\psi_{x}(end)$. Then the flux reduces to 
\begin{equation}
Flux = \frac{1}{2}[2\psi^{j}(end)-\psi^{j-1}(end)+2\phi^{j}(1)-\phi^{j-1}(1)].
\end{equation}
For example, at steady state $\psi^{j}(end)=\psi^{j-1}(end)=\psi^{j+1}(end)$ and so the flux is $Flux = \frac{1}{2}[\psi^{j}(end) +\phi^{j}(1)].$

Similarly for the 4th-order method,  
\begin{eqnarray}
\phi^{j+1}(end) &=& 2\phi^{j}(end) - \phi^{j-1}(end)+
\frac{{\Delta t}^2}{h_{1}^{2}\lambda^2(\lambda^2-1)}[-\phi^{j}(end-2)+ \nonumber \\
&&\lambda^4\phi^{j}(end-1)-2(\lambda^4-1)\phi^{j}(end)+
\lambda^4\psi^{j}(ghost1)-\psi^{j}(ghost2)] \nonumber
\end{eqnarray}
and 
\begin{eqnarray}
\psi^{j+1}(1) &=& 2\psi^{j}(1) - \psi^{j-1}(1)+
\frac{{\Delta t}^2}{h_{1}^{2}\lambda^2(\lambda^2-1)}[-\phi^{j}(ghost2)+ \nonumber \\
&&\lambda^4\phi^{j}(ghost1)-2(\lambda^4-1)\psi^{j}(1)+
\lambda^4\psi^{j}(2)-\psi^{j}(3)]\nonumber . 
\end{eqnarray}
For SP-SP hybrid process, and at the $j$th interface $\phi^{j}(ghost2)=\phi^{j}(end-2)$ , $\phi^{j}(ghost1)=\phi^{j}(end-1)$ , $\psi^{j}(ghost1)=\psi^{j}(2)$ and $\psi^{j}(ghost2)=\psi^{j}(3)$. We can write the central flux at the $jth$ interface as follows
\begin{eqnarray}
Flux &=& \frac{\phi^{j+1}(end)+\psi^{j+1}(1)}{2}  \nonumber \\ 
 &=& \frac{(2\phi^{j}(end) - \phi^{j-1}(end) + 2\psi^{j}(1) - \psi^{j-1}(1))}{2} + \nonumber \\
&& \frac{{\Delta t}^2}{2h_{1}^{2}\lambda^2(\lambda^2-1)}[-\phi^{j}(end-2)+ \nonumber \\ 
 &&\lambda^4\phi^{j}(end-1)-2(\lambda^4-1)\phi^{j}(end)+
\lambda^4\psi^{j}(ghost1)-\nonumber \\
&&\psi^{j}(ghost2)+\phi^{j}(ghost2)+\nonumber \\ 
&&\lambda^4\phi^{j}(ghost1)-2(\lambda^4-1)\psi^{j}(1)+
\lambda^4\psi^{j}(2)-\psi^{j}(3)].
\end{eqnarray}
 
If we use the central flux directly then
\begin{eqnarray}
\psi^{j+1}(end) &=& 2\psi^{j}(end) - \psi^{j-1}(end) + (\Delta t)^2D^2\psi^j , \nonumber \\ 
\phi^{j+1}(1) &=& 2\phi^{j}(1) - \phi^{j-1}(1) + (\Delta t)^2D^2\phi^j ,  \nonumber \\
\end{eqnarray}
where $D$ is the spectral differentiation matrix. 
\begin{eqnarray}
Flux &=&\frac{1}{2}[2\psi^{j}(end)-\psi^{j-1}(end)+2\phi^{j}(1)-\phi^{j-1}(1)]+\nonumber \\
      &&\frac{(\Delta t)^2}{2}[D^2\psi^j_{end} +D^2\phi_{1}^{j}].
\end{eqnarray}
If the function is not $C^{2}$ then $D^2\psi_{end}^{j} \ne D^2\phi_{1}^{j}$.
For the FD-SP domain the resolutions are usually different, i.e $\phi^{j}(ghost) \ne \phi^{j}(2)$ and $\psi^{j}(ghost) \ne \psi^{j}(end - 1)$. 
But we can expand $\phi^j$ and $\psi^j$ around the $2^{nd}$ and $(end -1)$-th point respectively.

\section{Boundary conditions}
For the boundary conditions at the boundaries $\partial \Omega$ of the whole domain $\Omega$, 
we use the simple outflow condition based on the assumption that the potential term at $\partial\Omega$ 
is negligible. That is, 
we use 
\begin{eqnarray}
  &&\psi_t - \psi_{r^*} = 0, \quad r^* \in \partial^- \Omega^1_s, \nonumber \\
  &&\psi_t + \psi_{r^*} = 0, \quad r^* \in \partial^+ \Omega^M_f. \nonumber 
\end{eqnarray}
For the first-order derivative in the outflow equations, we use the 2nd-order finite-difference method, Eq. (\ref{left2ndorder}) 
for the FD domain $\Omega^M_f$ and Eq. (\ref{right2ndorder2}) for the SP domain $\Omega^1_s$. 


\section{Stability analysis}

\vskip 0.2in
\setlength{\unitlength}{4cm}
\begin{picture}(1,1)
\put(1.25,1.25){\vector(-1,-1){0.2}}
\put(0.75,1){\circle{0.09}}
\put(0.95,1){\circle{0.09}}
\put(1.1,1){\circle{0.09}}

%
\put(1.1,1){\circle*{0.05}}
\put(1.25,1){\circle*{0.09}}
\put(1.45,1){\circle*{0.09}}

\put(1.27,1.25){$\phi^{n+1}_N, \psi^{n+1}_1$}

\put(-0.1,1){\line(5,0){2.3}}
\put(0,0.5){$\begin{array}{clrl}
   \bigcirc & \mbox{SP1,} & {\bullet}  &\mbox{SP2}\end{array} $} 
\put(-0.1,0.3){Diagram 3: SP - SP patching. Grids at the SP and SP interface.
}
\label{diagram3}
\end{picture}
\subsection{SP-SP stability analysis}
For stability analysis we consider SP-SP domain for example. This analysis is for two domains. Let us consider the two spectral domains denoted by SP1 and SP2. Let $\phi$ be the solution in SP1 and $\psi$ be the solution in SP2 of the Zerrilli equation without the potential and singular source terms.
Then we have
\begin{eqnarray}
\phi^{n+1}_{j} = 2\phi^{n}_{j}-\phi^{n-1}_{j}+(\Delta t)^2\tilde{D_{1}^2}\phi^{n}_{j} , \: j = 1,\cdots, N-1,\nonumber \\
\phi^{n+1}_N = 2\phi^{n}_N-\phi^{n-1}_N+\frac{(\Delta t)^2}{2h^2}[\phi^{n}_{N-1}-2\phi^{n}_{N}+\psi^{n}_{1}], \nonumber \\
\psi^{n+1}_0 = 2\psi^{n}_0-\psi^{n-1}_0+\frac{(\Delta t)^2}{2h^2}[\phi^{n}_{N-1}-2\psi^{n}_{0}+\psi^{n}_{1}], \nonumber \\
\psi^{n+1}_{j} = 2\psi^{n}_{j}-\psi^{n-1}_{j}+(\Delta t)^2\tilde{D_{2}^2}\psi^{n}_{j} , \: j = 1,\cdots, N-1.\nonumber
\end{eqnarray}
We assume that there is no boundary effect. Also we consider the equal length intervals and so $D_{1} = D_{2} = \frac{2}{L}D$ where $D$ is the original Chebyshev differential matrix and $L$ is the length of the spectral sub-domains. $\tilde{D_{1}^2}$ is the sub-matrix of $D_{1}^2$ without the last and first rows and the first column and $\tilde{D_{2}^2}$ is the sub-matrix of $D_{2}^2$ without last row, first row and first column.\\
Collecting all the above equations in matrix form yields,
$$\left[\begin{array}{c} \phi^{n+1}_{1} \\ \phi^{n+1}_{2} \\ \vdots \\ \phi^{n+1}_{N}\\\psi^{n+1}_{0}\\ \vdots \\ \psi^{n+1}_{N-1} \end{array}\right]
 = 2\left[\begin{array}{c} \phi^{n}_{1} \\ \phi^{n}_{2} \\ \vdots \\ \phi^{n}_{N}\\\psi^{n}_{0}\\ \vdots \\ \psi^{n}_{N-1} \end{array}\right]
  - \left[\begin{array}{c} \phi^{n-1}_{1} \\ \phi^{n-1}_{2} \\ \vdots \\ \phi^{n-1}_{N}\\\psi^{n-1}_{0}\\ \vdots \\ \psi^{n-1}_{N-1} \end{array}\right]$$
    $$+  (\Delta t)^2\left[\begin{array}{cccccccccccc}\: & \: & \: & \: & \:  &\: &\: &\: &\: &\: \\ \: &\: & (\tilde{D_{1}^{2}})_{N-1, N} & \: & \: & \:  &\: &\begin{LARGE} \mathbb{N} \end{LARGE} & \: &\: \\ \: & \: & \: & \: & \:  &\: &\: &\: &\: &\: \\ 0 & 0 & \cdots  & \frac{1}{2h^2} & -\frac{1}{h^2} & 0 & \frac{1}{2h^2} & \cdots & 0  &0 \\  0  & 0 &\cdots & \frac{1}{2h^2} & 0  & -\frac{1}{h^2} & \frac{1}{2h^2} &\cdots &0 &0 \\ \: & \: & \: & \: & \:  &\: &\: &\: &\: &\:\\ \: &\: & \begin{LARGE} \mathbb{N} \end{LARGE} & \: & \: & \:  &\: &(\tilde{D_{2}^{2}})_{N-1, N} & \: &\: \\\: & \: & \: & \: & \:  &\: &\: &\: &\: &\: \end{array}\right]
    \left[\begin{array}{c} \phi^{n}_{1} \\ \phi^{n}_{2} \\ \vdots \\ \phi^{n}_{N}\\\psi^{n}_{0}\\ \vdots \\ \psi^{n}_{N-1} \end{array}\right],$$ 
where $\begin{LARGE}\mathbb{N}\end{LARGE}$ is the $N-1 \times (N)$  null matrix.
The previous matrix equation can be written in compact form $$W^{n+1} = 2W^{n}-W^{n-1}+\tilde{D}W^{n},$$ or \begin{eqnarray}W^{n+1} = (2I_{2N}+\tilde{D})W^{n}-W^{n-1}, \\ W^{n} = I_{2N}W^{n}.\end{eqnarray} Eqs $23$ and $24$ can be written in matrix form
$$\left[\begin{array}{c}W^{n+1}\\W^{n}\end{array}\right]=\left[\begin{array}{cc}2I_{2N}+\tilde{D} & -I_{2N} \\I_{2N} & 0 \end{array}\right] \left[\begin{array}{c}W^{n}\\W^{n-1}\end{array}\right].$$ Let $$A = \left[\begin{array}{cc}2I_{2N}+\tilde{D} & -I_{2N} \\I_{2N} & 0 \end{array}\right].$$ For the matrix stability, we have the spectral radius of $A$ less than $1$. i.e $\rho(A) < 1$.\\ \\
To check the stability we consider two spectral domains $[-1, 1]$ and $[1, 3]$. For these two domains $D_{1}$ and $D_{2}$ must be equal. Let $h_{1}$ be the minimum grid spacing in the left domain and similarly $h_{2}$ for the right domain. Here we have $h_{1} = h_{2} = h$. We found the following results for $N$(number of grid points in each domain) and $\Delta t$(time step) for which $\rho(A) < 1$.

\begin{table}[ht]
\tbl{Comparison of number of grid points $(N)$, time step $(\Delta t)$, minimum grid spacing $(h)$.}
{\begin{tabular}{@{}ccc@{}} \toprule
 \qquad \qquad $N$ \qquad \qquad &  \qquad \qquad $\Delta t$ \qquad \qquad & \qquad \qquad $h$ \qquad \qquad \\
 \\ \colrule
 \hphantom{0}16 & \hphantom{0}$10^{-5}$ & 7.810 $\times 10^{-3}$ \\ \hphantom{0}32 & \hphantom{0}$10^{-6}$ & 1.953 $\times 10^{-3}$\\
 \hphantom{0}64 & \hphantom{0}$10^{-6}$ & 4.883 $\times 10^{-4}$\\ \hphantom{0}128 & \hphantom{0}$10^{-7}$ & 1.221 $\times 10^{-4}$\\ \hphantom{0}256 & \hphantom{0}$10^{-7}$ & 3.052 $\times 10^{-5}$\\  \hphantom{0}512 & \hphantom{0}$10^{-8}$ & 7.629 $\times 10^{-6}$\\ \botrule
\end{tabular} \label{ta1}}
\end{table}


\subsection{SP-FD stabilty analysis}

\vskip 0.5in
\setlength{\unitlength}{4cm}
\begin{picture}(1,1)
\put(1.25,1.25){\vector(-1,-1){0.2}}
\put(0.75,1){\circle{0.09}}
\put(0.95,1){\circle{0.09}}
\put(1.1,1){\circle{0.09}}

%
\put(1.1,1){\circle*{0.05}}
\put(1.25,1){\circle*{0.09}}
\put(1.4,1){\circle*{0.09}}

\put(1.27,1.25){$\phi^{n+1}_N, \psi^{n+1}_0$}
\put(1.15,0.9){$h$}
\put(1.30,0.9){$h$}
\put(0.82,0.9){$h_{1}$}
\put(1.0,0.9){$h$}
\put(-0.1,1){\line(5,0){2.3}}
\put(0,0.5){$\begin{array}{clrl}
\bigcirc & \mbox{SP,} & {\bullet}  &\mbox{FD}\end{array} $} 
\put(-0.1,0.3){Diagram 4: SP - FD patching. Grids at the SP and FD interface.
}
\label{diagram4}
\end{picture}

At the SP-FD interface the grid resolutions between the adjacent sub-domains across the domain interface are different. So the grid distribution is non-uniform. The stable interface conditions derived for the uniform grid system are not enough and we need some conditions with which the spatial non-uniformity can be addressed properly [see 26 and references therein]. Let $N_{1}$ be the number of grid points in each SP sub-domain and $N_{2}$ be the number of grid points in the FD domain. Also assume $L_{1}$ be the length  of each SP sub-domain and $L_{2}$ be the length of the FD sub-domain. Then for stability we must have \cite{Don_Gottlieb}
\begin{equation} \frac{L_{1}}{N_{1}^{2}}=\frac{L_{2}}{N_{2}}. \end{equation}

Consider  one spectral domain (SP1) and one FD-domain (FD1). Let $\phi$ be the solution in SP1 and $\psi$ be the solution in FD1 of the simple wave equation without any potential term. Let number of grid points in SP1 and FD1 be $N_{1}$ and $N_{2}$ respectively which satisfy eqn (25). We assume that there is no boundary effect.
Then from spectral collocation  and $2$nd order FD-method, we have
\begin{eqnarray}
\phi^{n+1}_{j} = 2\phi^{n}_{j}-\phi^{n-1}_{j}+(\Delta t)^2\tilde{D_{1}^2}\phi^{n}_{j} , \: j = 1,\cdots, N_{1}-1,\nonumber \\
\phi^{n+1}_{N_{1}} = 2\phi^{n}_{N_{1}}-\phi^{n-1}_{N_{1}}+\frac{(\Delta t)^2}{2h^2}[\phi^{n}_{N_{1}-1}-2\phi^{n}_{N_1}+\psi^{n}_{1}], \nonumber \\
\psi^{n+1}_0 = 2\psi^{n}_0-\psi^{n-1}_0+\frac{(\Delta t)^2}{2h^2}[\phi^{n}_{N_{1}-1}-2\psi^{n}_{0}+\psi^{n}_{1}], \nonumber \\
\psi^{n+1}_{j} = 2\psi^{n}_{j}-\psi^{n-1}_{j}+(\Delta t)^2\tilde{D_{2}}\psi^{n}_{j} , \: j = 1,\cdots, N_{2}-1.\nonumber
\end{eqnarray}
Where $D_{1} = \frac{2}{L}D$ where $D$ is the original Chebyshev matrix, $\tilde{D_{1}^2}$ is the sub-matrix of $D_{1}^2$ without the first and last rows and the last column and $D_{2}$ is the $2$nd-order FD-differentiation matrix, which is written by
$$D_{2} = \frac{1}{2h^2} \left[\begin{array}{ccccccc} 1 & -2 & 1 & \: & \: & \: & \: \\ \: & 1 & -2 & 1 & \: & \: & \: \\ \: & \: & . & . & . & \: & \:  \\ \: & \: & \: & . & . & . & \: \\ \: & \: & \: & \: & 1 & -2 & 1  \end{array}\right].$$ 
\\
and $\tilde{D_{2}}$ is the sub-matrix of $D_{2}$ without the first and last rows and the first column.
Collecting all the above equations in matrix form yields,
$$
\left[\begin{array}{c} 
\phi^{n+1}_{1} \\ \phi^{n+1}_{2} \\ \vdots \\ \phi^{n+1}_{N_{1}}\\\psi^{n+1}_{0}\\ \vdots \\ \psi^{n+1}_{N_{2}-1} \end{array}\right]
 = 2\left[\begin{array}{c} \phi^{n}_{1} \\ \phi^{n}_{2} \\ \vdots \\ \phi^{n}_{N_{1}}\\\psi^{n}_{0}\\ \vdots \\ \psi^{n}_{N_{2}-1} \end{array}\right]
  - \left[\begin{array}{c} \phi^{n-1}_{1} \\ \phi^{n-1}_{2} \\ \vdots \\ \phi^{n-1}_{N_{1}}\\\psi^{n-1}_{0}\\ \vdots \\ \psi^{n-1}_{N_{2}-1} \end{array}\right] $$
   $$ +  (\Delta t)^2\left[\begin{array}{cccccccccccc}\: & \: & \: & \: & \:  &\: &\: &\: &\: &\: \\ \: &\: & (\tilde{D_{1}^{2}})_{N_{1}-1, N_{1}} & \: & \: & \:  &\: &\begin{LARGE}\mathbb{N}_{1} \end{LARGE} & \: &\: \\ \: & \: & \: & \: & \:  &\: &\: &\: &\: &\: \\ 0 & 0 & \cdots  & \frac{1}{2h^2} & -\frac{1}{h^2} & 0 & \frac{1}{2h^2} & \cdots & 0  &0 \\  0  & 0 &\cdots & \frac{1}{2h^2} & 0  & -\frac{1}{h^2} & \frac{1}{2h^2} &\cdots &0 &0 \\ \: & \: & \: & \: & \:  &\: &\: &\: &\: &\:\\ \: &\: & \begin{LARGE} \mathbb{N}_{2} \end{LARGE} & \: & \: & \:  &\: &(\tilde{D_{2}})_{N_{2}-1, N_{2}} & \: &\: \\\: & \: & \: & \: & \:  &\: &\: &\: &\: &\: \end{array}\right]
   \left[\begin{array}{c} \phi^{n}_{1} \\ \phi^{n}_{2} \\ \vdots \\ \phi^{n}_{N_{1}}\\\psi^{n}_{0}\\ \vdots \\ \psi^{n}_{N_{2}-1} 
\end{array}\right],
$$ 
Where $\begin{LARGE}\mathbb{N}_{1}\end{LARGE}$ is the $N_{1}-1 \times (N_{2})$  null matrix, and $\begin{LARGE}\mathbb{N}_{2}\end{LARGE}$ is the $N_{2}-1 \times (N_{1})$  null matrix.
The previous matrix equation can be written in the following compact form $$W^{n+1} = 2W^{n}-W^{n-1}+\tilde{D}W^{n},$$ or \begin{eqnarray}W^{n+1} = (2\tilde{I}+\tilde{D})W^{n}-W^{n-1}, \\ W^{n} = \tilde{I}W^{n}.\end{eqnarray} Eqs $26$ and $27$ can be written in matrix form
$$\left[\begin{array}{c}W^{n+1}\\W^{n}\end{array}\right]=\left[\begin{array}{cc}2\tilde{I}+\tilde{D} & -\tilde{I} \\\tilde{I} & 0 \end{array}\right] \left[\begin{array}{c}W^{n}\\W^{n-1}\end{array}\right].$$ Let $$\tilde{A} = \left[\begin{array}{cc}2\tilde{I}+\tilde{D} & -\tilde{I} \\\tilde{I} & 0 \end{array}\right].$$ Where $\tilde{I}$ is the unit matrix of order $N_{1}+N_{2}$. For the matrix stability, we must have the spectral radius of $\tilde{A}$ less than $1$. i.e $\rho(\tilde{A}) < 1$.\\ \\
To check the stability we consider the SP-domain $[-1 \quad 1]$ and FD-domain $[1 \quad 2]$. We choose $N_1$ and $N_2$ in such a way that they must satisfy Eq. (25). We found the following results for $N_1$(number of grid points in SP-domain), $N_2$ (Number of grid points in FD-domain) and $\Delta t$(time step) for which $\rho(\tilde{A}) < 1$. Some examples are given in the following table:

\begin{table}[ht]
\tbl{Comparison of number of grid points in SP domain $(N_{1})$, number of grid points on FD domain $(N_{2})$ and time step $\Delta t$.}
{\begin{tabular}{@{}ccc@{}} \toprule
 \qquad \qquad $N_{1}$ \qquad \qquad &  \qquad \qquad $N_{2}$ \qquad \qquad & \qquad \qquad $\Delta t$ \qquad \qquad \\
 \\ \colrule
 \hphantom{0}$8$ & \hphantom{0}$64$ &  $ 10^{-5}$ \\ \hphantom{0}$16$ & \hphantom{0}$256$ & $10^{-6}$\\
 \hphantom{0}$32$ & \hphantom{0}$1024$ & $ 10^{-7}$\\ \hphantom{0}$48$ & \hphantom{0}$2304$ & $ 10^{-8}$\\  \botrule
\end{tabular} \label{ta1}}
\end{table}

The table implies that the more uniformity of the grid spacing across the SP and FD interface is achieved the larger value of time step could be used for stability. For our numerical experiments in the next section, we choose the geometric parameters for the sub-domains so that the grid resolution is uniform in the small neighborhood across the SP-FD interface. Here note that for the numerical experiments in this paper, we choose the geometric setting so that we have about $\Delta t \sim 10^{-3}$.

\section{Numerical results}

For the numerical experiments, the computational domain is split into two main sub-domains, i.e. the SP and FD domains. The $\delta$-function is located in the SP sub-domain for all time (see Diagram 5). The SP sub-domain is also divided into multiple smaller SP sub-domains and each sub-domain communicates with its adjacent sub-domains by the interface condition described in Section 5. The last SP sub-domain communicates with the boundary FD sub-domain. 

\vskip 1.2in
\setlength{\unitlength}{4cm}
\begin{picture}(1,1)
\put(0.15,1.22){\vector(-1,-1){0.2}}
\put(0.15,1.25){ SP sub-domains}
\put(0.80,1.22){\vector(1,-1){0.2}}

\put(1.38,1.22){\vector(-1,-1){0.2}}
\put(1.30,1.25){ FD sub-domain}
\put(1.90,1.22){\vector(1,-1){0.2}}

\put(0.70,1){\line(0,1){0.04}}
\put(0.9,1){\line(0,1){0.04}}
\put(1.1,1){\line(0,1){0.04}}
\put(0.5,1){\line(0,1){0.04}}
\put(0.3,1){\line(0,1){0.04}}
\put(0.1,1){\line(0,1){0.04}}
\put(-0.1,1){\line(0,1){0.04}}

\put(2.2,1){\line(0,1){0.04}}

\put(-0.1,1){\line(5,0){2.3}}

\put(-0.1,0.6){Diagram 5: SP - SP .... - SP - FD subdomains.}
\label{diagram5}
\end{picture}

Since the spectral method is a global method, the boundary effects spread instantly throughout the domain. Thus, the SP method suffers from the boundary effects unless the range of the domain is large enough. However, the multiple SP sub-domains with the FD boundary domain help avoid the unphysical oscillations and reduce the computational cost.

To determine the number of sub-domains and the order of the interpolating polynomial in each sub-domain, two aspects are considered. The first aspect is the resolution for the singular source term and the second one is the grid uniformity across the SP and FD domain interface. As explained in the previous section, the non-uniformity of the grid resolution near the SP and FD interface makes the CFL condition strict. Considering  these aspects, 
we truncate the domain to $[-300, 387.5]$ in $r^*$. The SP sub-domain is $[-300, 20]$ and the FD sub-domain is $[20, 387.5]$. The multi-domain SP method reduces the computational time significantly by reducing the size of the system but still achieves the desired accuracy. The typical setup  in this work is that the number of sub-domains in the SP domain is $(N)=40$ and each domain has the interpolating order of $(n)=48$. The time-step is relatively large such as  $\sim 10^{-3}$ and the Gaussian $\delta$-function has $\sigma = 20$. The waveform is collected at various values of $r^{*}$.



\subsection{Gaussian $\delta$-function}
\begin{figure}[h]
	\centering
		\includegraphics[width=1.1\textwidth]{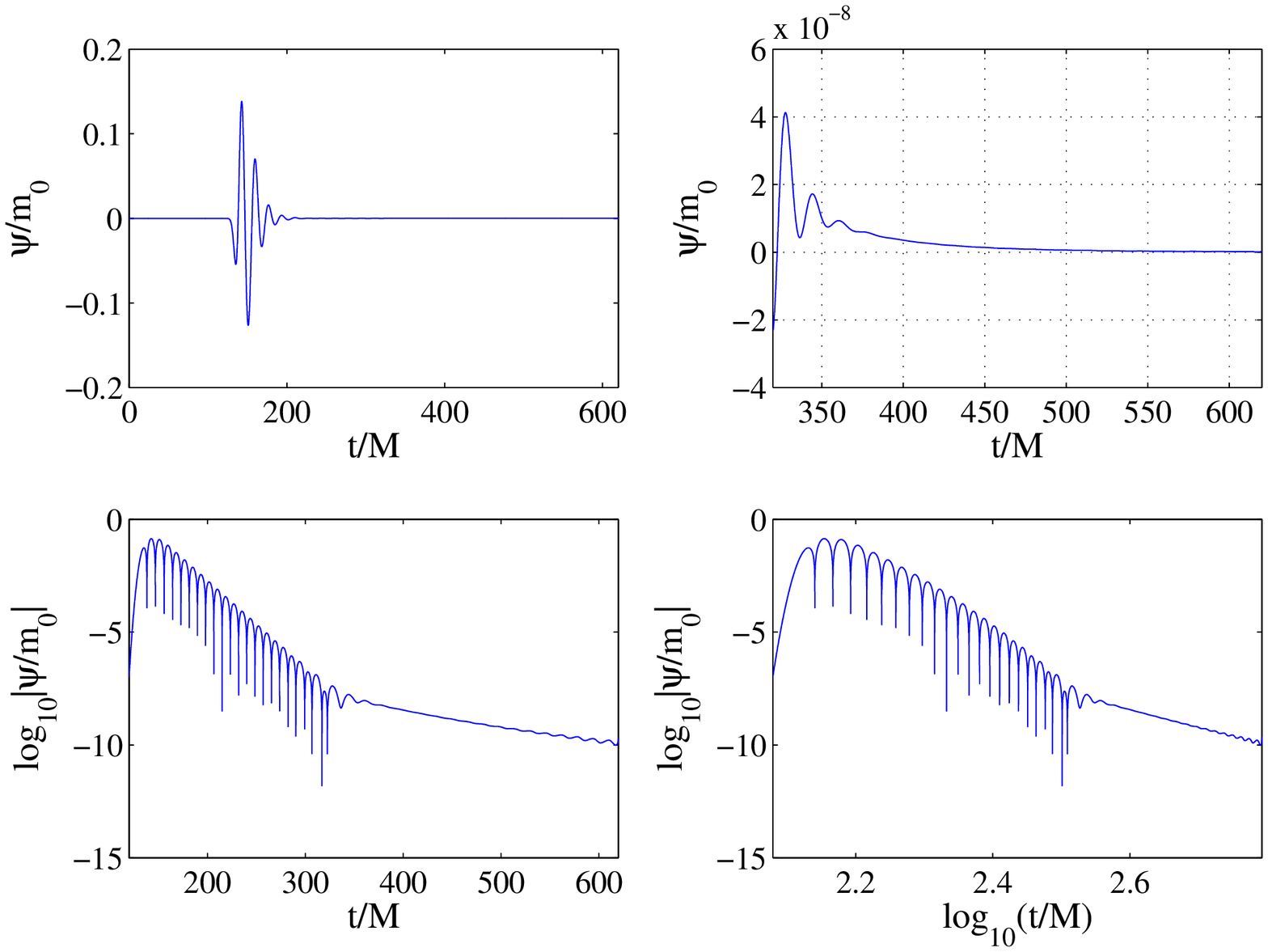}
		\caption{The QMN andthe  power-law decay with the Gaussian model with $\sigma = 20$. $R_{e}=-300$, $R_{\infty} = 387.5$, $r_{0} = 1.5(2M)$.}
	 \label{fig:Gauusian1}
\end{figure}
First we consider the case  with the Gaussian model. The first result is in Figure 1. The waveform is collected at $r^* = 137.6$. 
The left figure in the top panel shows the ringdown profile which starts around the time $t = 150$.  The right figure in the top panel shows the power-law decay. The two figures in the lower panel show the two distinct phases. For the time approximately $t = 150 \sim 300$ the solution decays exponentially and it is oscillatory. This phase is the Quasi-Normal-Mode (QNM) ringing phase.  After this phase the power-law decay starts. According to the seminal work by Richard Price \cite{R_Price}, the observer measures the late time perturbation field to drop off as an inverse power law of time, specifically as $t^{-n}$. In the case of Schwarzschild black hole, $n = 2l+3$, where $l$ is the multipole moment of the perturbation field. In the context of our present work \cite{Jung_Khanna_Nagle}, $l = 2$. The right figure in lower panel shows the power-law decay in logarithmic scale both in the horizontal and vertical axes. In this phase, the slope of the decay profile is about  $\sim t^{-6.7}$. The slopes of the power-law decay for various $\sigma$ and different resolutions, are almost same and $\sim -7$. This agrees well with the theoretical result~\cite{R_Price}: $\sim t^{-2l-3}$.

\begin{figure}[h]
	\centering
		\includegraphics[width=1.1\textwidth]{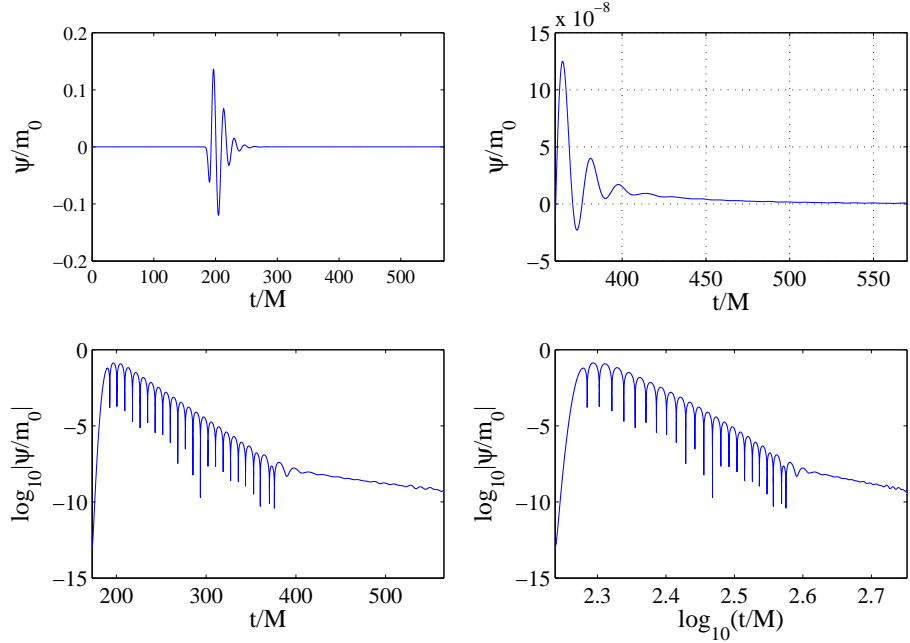}
		\caption{ The QMN and the power-law decay for Gaussian model with $\sigma = 10$. $R_{e}=-300$, $R_{\infty} = 387.5$, $r_{0} = 1.5(2M)$. }
	 \label{fig:Gauusian2}
\end{figure}
In Figure 2,  we use the Gaussian $\delta$-function with $\sigma = 10$. 
The waveform is collected  at $r^{*}= 137.6$.
The top panel shows the ringdown profile and lower panel shows the QNM and the power-law decay. 
The similar QNM and the power-law decay are observed. Also we find that the same number of oscillations are observed in the QNM regime as in Figure 1. 

\begin{figure}[h]
	\centering
		\includegraphics[width=1.1\textwidth]{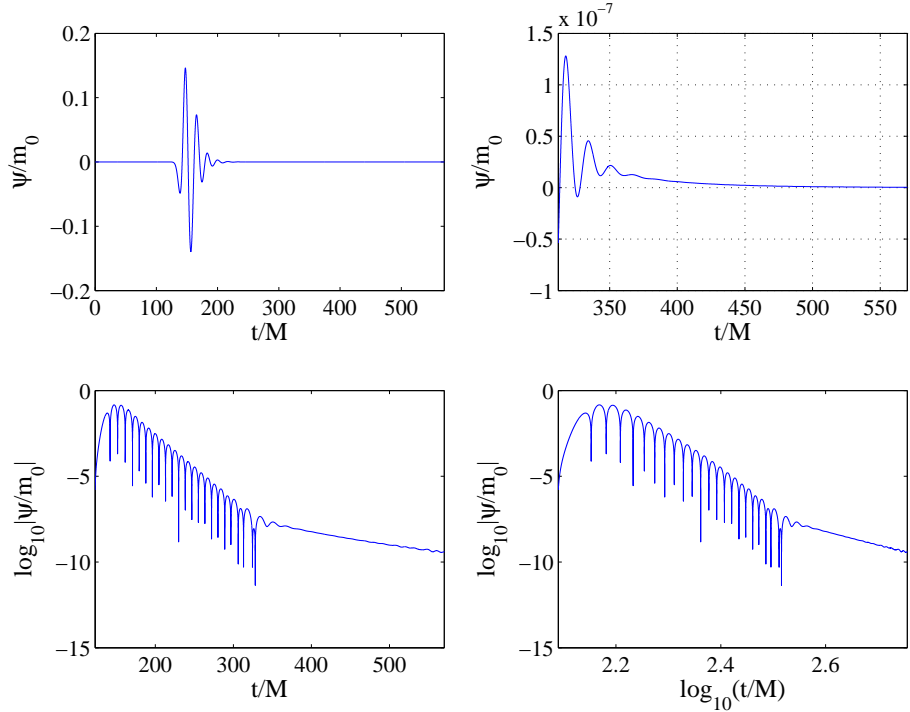}
		\caption{ The QMN and the power-law decay for the Gaussian model with $\sigma = 40$. $R_{e}=-300$, $R_{\infty} = 387.5$, $r_{0} = 1.5(2M)$.}
	 \label{fig:Gauusian3}
\end{figure} 
In Figure 3, we use the Gaussian model with $\sigma = 40$. The waveform is collected  at $r_{*}= 137.6$. Note that the higher value of $\sigma$ is used for this figure and the Gaussian $\delta$-function is much smoother than the previous two cases. We, however, found that the similar solution with the expected QNM profile and power-law decay profile was obtained. 

\begin{figure}[h]
		\includegraphics[width=1.1\textwidth]{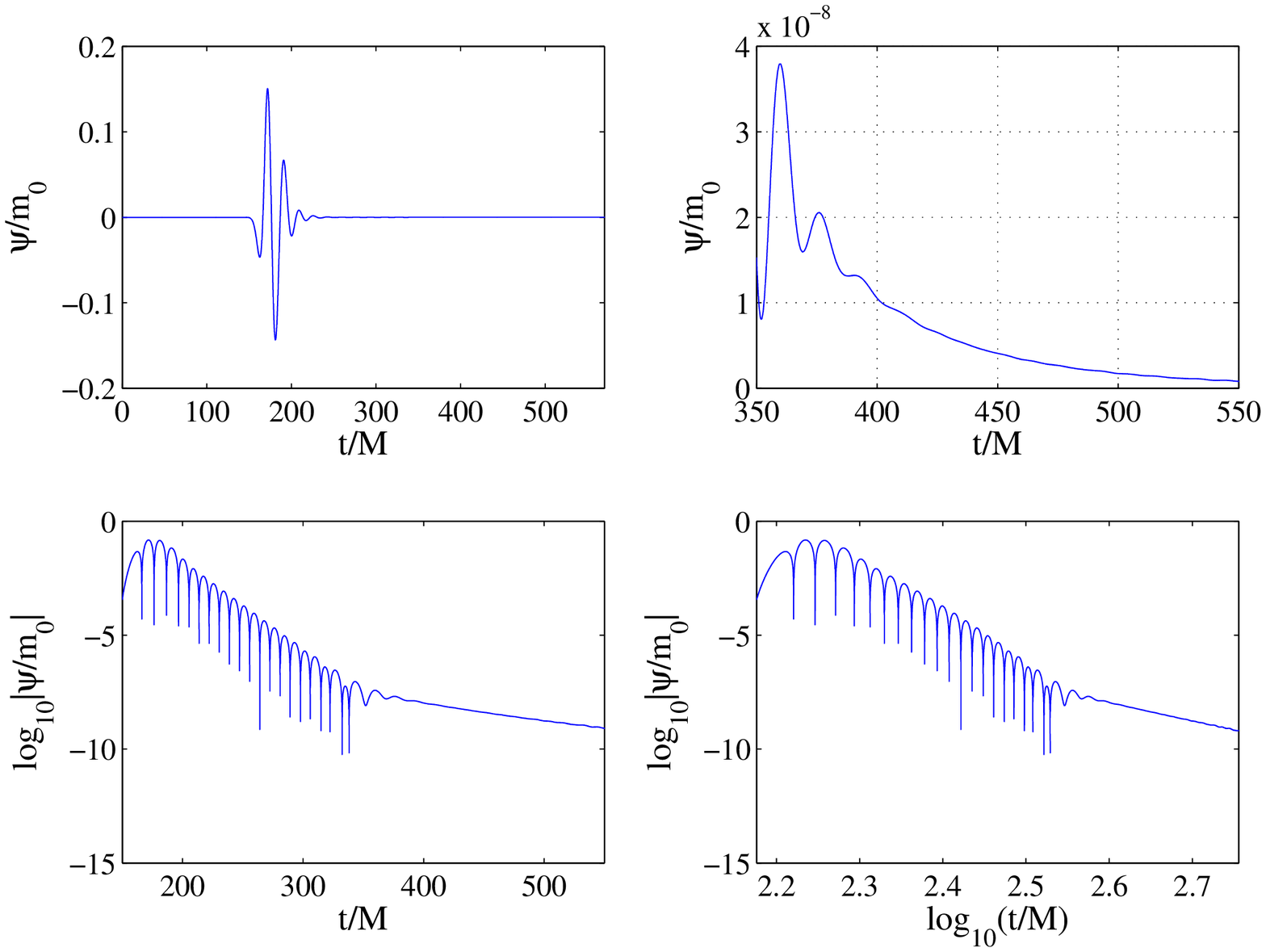}
		\caption{ The QMN and the power law decay for the Gaussian model with $\sigma = 50$. $R_{e}=-300$, $R_{\infty} = 387.5$, $r_{0} = 1.5(2M)$. }
	 \label{fig:Gauusian4}
\end{figure}

In Figure 4, we use the Gaussian $\delta$ model with $\sigma = 50$. The waveform was collected at $r_{*} = 137.6$. With this high value of $\sigma$, the Gaussian $\delta$-function is even smoother. This high level of smoothness of the singular source term made it unnecessary to use the spectral filter for the solution. For the previous figures, we used the spectral filter to regularize the solution due to the possible Gibbs oscillations. Without the filter operation, the computational time was reduced. This suggests that the desired decaying profile can be obtained by having the parameters in a clever way. For example, in Figure 5, we show the polynomial order $n$ that we used for figures with different values of $\sigma$ to obtain the desired QNM and the power-law decay.  The parameter values in the figure are 
$(n, \sigma) = (96, 10)$,  $(48, 20)$, $(32, 40)$, $(24, 50)$. In our current research we didn't attempt, but it would be an interesting study to investigate the optimal configuration of these parameters.

\begin{figure}[h]
	\centering
		\includegraphics[width=0.6\textwidth]{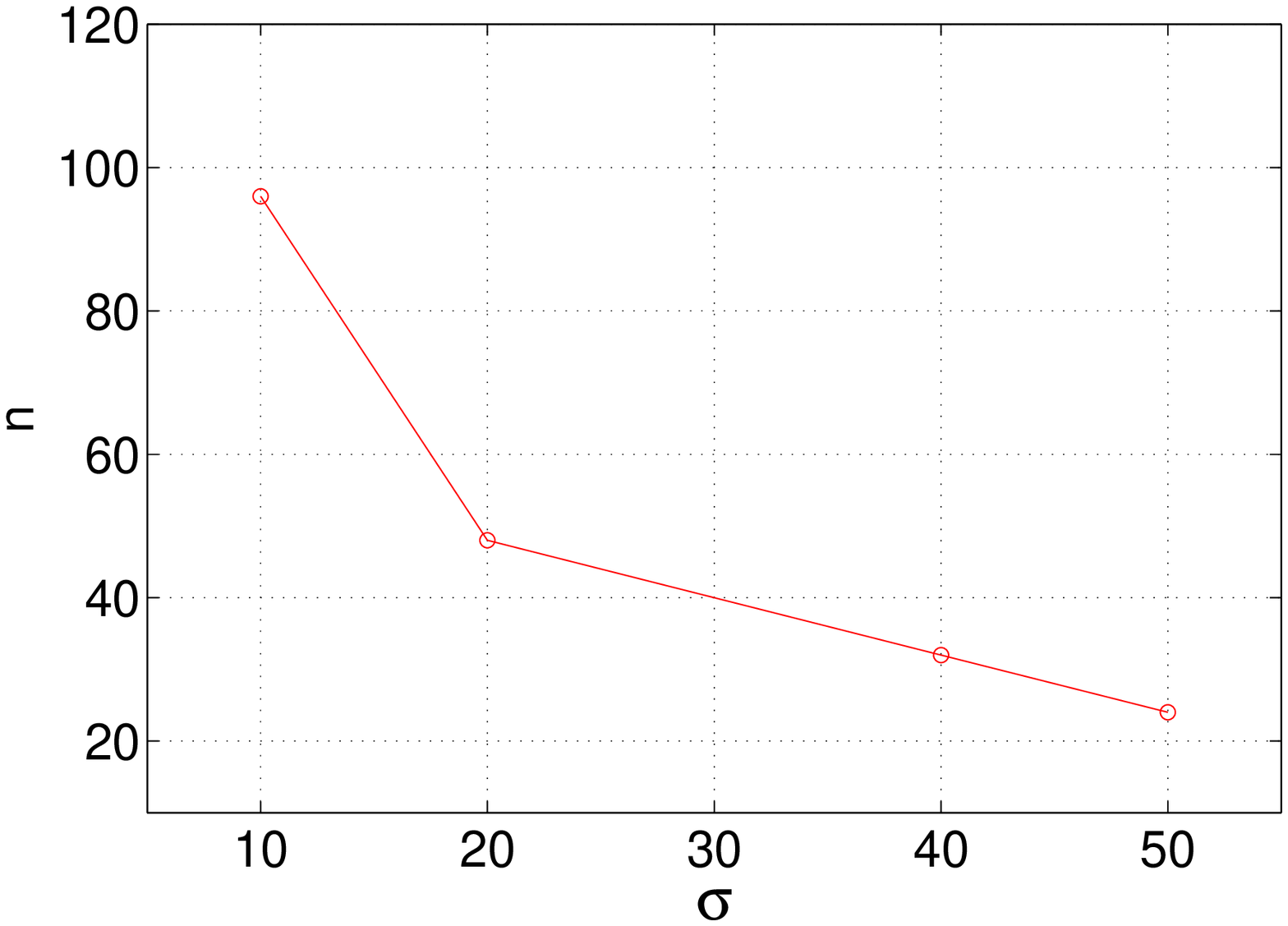}
		\caption{The polynomial order $n$ in each SP sub-domain with the value of $\sigma$ for the Gaussian model. 
		$N = 40$.}
	 \label{figure5}
\end{figure}


\subsection{Discrete $\delta$-function}
We repeated the above experiment with the discrete $\delta$-function. By definition,  the discrete $\delta$-function is localized and its shape changes with time because the spectral grid spacing is non-uniform.

\begin{figure}[h]
	\centering
		\includegraphics[width=1.1\textwidth]{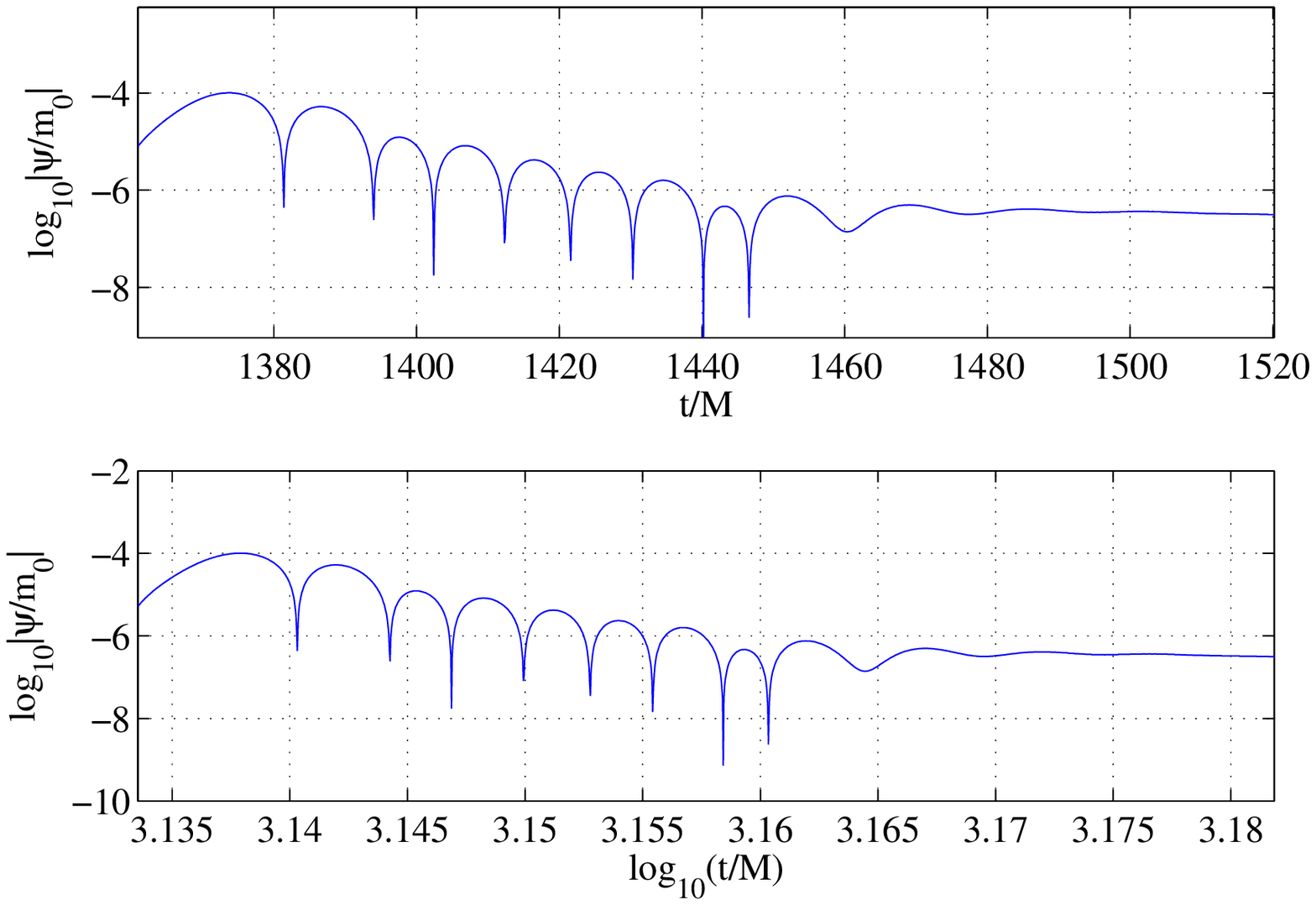}
		\caption{  The QNM and the power-law decay for the discrete $\delta$-function. $R_{e}=-300$, $R_{\infty} = 500$, $r_{0} = 3(2M)$.}
	 \label{fig:Discrete1}
\end{figure}
\begin{figure}[h]
	\centering
		\includegraphics[width=1.1\textwidth]{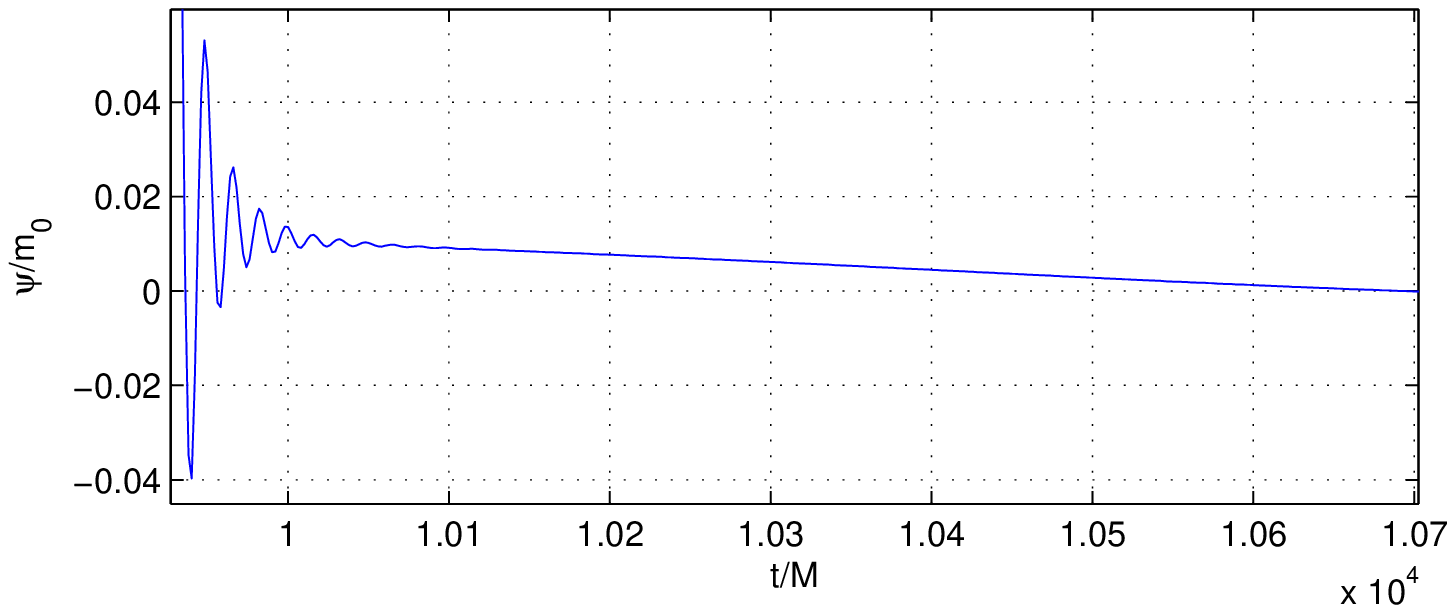}
		\caption{ The last few QNM modes and  the long power-law tail for the discrete $\delta$- function. $R_{e}=-30$, $R_{\infty} = 1970$, $r_{0} = 200(2M)$.}
	 \label{fig:Discrete2}
\end{figure}

The QNM and the power-law decay profiles are depicted in Fig. 6. But the slope of the power-law tail does not match with that for the Gaussian model. 
For example, in Fig. 6 we considered the case that $r_{e} = -300$, $r_{\infty} = 500$ and the interface between the SP and FD domains is at $r_{*}= 150$. We use $40$ SP sub-domains with $n = 48$. The waveform was collected at $r_{*}=250$. The decay rate is slower than expected. The estimated slope is $\sim -2$ while the slope with the Gaussian model was close to $-7$. Although we increase the resolution, $n$, no significant improvement was observed.
Fig. 7 shows the QNM and the power-law tail with the discrete $\delta$-function. Here we used the very far outer boundary $r_{\infty}=1970$. 
The slope is still about $-2$. It seems that the discrete $\delta$-function is good for the FD domain with the uniform grid but it is not ideal for the non-uniform grid  with the SP domain. 

\section{Conclusion}
In this work,  the inhomogeneous Zerilli equation was solved numerically in time-domain, for which we developed a multi-domain hybrid method based on the Chebyshev spectral collocation method and the $4$th-order explicit finite-difference method. Using the developed method, we computed the waveforms for head on collisions of black holes in one dimension. 
The in-falling black hole is modeled as a point source and consequently singular source terms appear in the governing equation. For the approximation of singular source terms, we used both the Gaussian $\delta$-function and the discrete $\delta$-function. For the stable and accurate approximation, we derived the interface conditions between the spectral and spectral domains and the spectral and finite-difference domains. The main approach introduced in this work is the use of the finite-difference domain as the boundary domain. Without the finite-difference domain as the boundary domain, the multi-domain composed of only the spectral sub-domains does not yield the proper power-law decay profile unless the range of the computational domain is very large. Using the multi-domain approach with the finite-difference boundary domain method, we could obtain the proper power-law decay profile with a relatively small computational cost. That is, the CFL condition is much relaxed and the location of the outer boundary of the computational domain is not afar. Numerical results show that the hybrid method obtains a proper quasi-normal and power-law decay with the Gaussian $\delta$-function approximation. Interestingly, even with a large value of $\sigma$, the proper power-law decay was observed. With the large value of $\sigma$, the spectral filtering operation was not necessary, so the computational time was much reduced. The hybrid method with the discrete $\delta$-function approximation, however, did not yield the proper power-law decay. The current study only considered the multi-domain spectral and finite-difference method with the $\delta$-function residing in the spectral domain. We will investigate the optimal configuration of the multi-domain computational domain in our future research.

\end{document}